# Ultrafast Charge Migration Dynamics in Enol Keto Tautomerization Monitored with a Local Soft-X-Ray Probe

Micheline B. Soley,[+,*,[a,b]] Pablo E. Videla,[+,*,[a,c]] Erik T. J. Nibbering,[d] and Victor S. Batista*[a-c]

[a] Dr. M. B. Soley,[+] Dr. P. E. Videla,[+] Prof. V. S. Batista*
Department of Chemistry, Yale University
P.O. Box 208107, New Haven, CT 06520-8107 (USA)
E-mail: micheline.soley@yale.edu
pablo.videla@yale.edu
victor.batista@yale.edu
[b] Dr. M. B. Soley, Prof. V. S. Batista
Yale Quantum Institute, Yale University
P.O. Box 27394, West Haven, CT 06520-8263 (USA)
[c] Dr. P. E. Videla, Prof. V. S. Batista
Energy Sciences Institute, Yale University
P. O. Box 27394, West Haven, CT 06516-7394 (USA)
[d] Dr. E. T. J. Nibbering
Max Born Institute for Nonlinear Optics and
Short Pulse Spectroscopy
Max Born Strasse 2A, 12489 Berlin (Germany)
[+] These authors contributed equally to this work.

**Abstract:** Proton-coupled electron transfer (PCET) is the underlying mechanism governing important reactions ranging from water splitting in photosynthesis to oxygen reduction in hydrogen fuel cells. The interplay of proton and electronic charge distribution motions can vary from sequential to concerted schemes, with elementary steps occurring on ultrafast time scales. We demonstrate with a simulation study that femtosecond soft-X-ray spectroscopy provides key insight into the PCET mechanism of a photoinduced intramolecular enol* → keto* tautomerization reaction. A full quantum treatment of electronic and nuclear dynamics of 2-(2-hydroxyphenyl-)benzothiazole upon electronic excitation reveals how spectral signatures of local excitations from core to frontier orbitals display the distinct stages of charge migration for the H atom, donating, and accepting sites. Our findings indicate UV/X-ray pump-probe spectroscopy provides a unique way to probe ultrafast electronic structure rearrangements in photoinduced chemical reactions essential to understanding the mechanism of PCET.

## Introduction

Understanding the dynamics of electronic rearrangements associated with proton transfer reactions remains a subject of great fundamental interest. Charge migrations are governed by the underlying microscopic mechanism generally known as proton-coupled electron transfer (PCET),[1] where the time scales for motions of electronic degrees of freedom (electronic charge distributions) and of nuclear degrees of freedom (of the transferring proton, and of hydrogen bond deformation modes) may be different or identical. PCET will then involve sequential or concerted reaction pathways, respectively. Examples where PCET plays a key role include water splitting by Photosystem II, nitrogen fixation, oxygen reduction in biocatalysis and in hydrogen fuel cell technologies.[2]

Significant molecular rearrangements induced by photoinduced proton transfer along pre-existing hydrogen bonds have been reported for many years, going back at least to the pioneering work of Weller on derivatives of salicylic acid monitored by UV-vis spectroscopy.[3] Since then, numerous studies have focused on understanding the process of photoinduced proton transfer in a variety of molecular systems, including hydroxyflavones, salicylaldehydes, 2-(2'-hydroxyphenyl)-benzothiazole derivatives, and related molecules.[4] Nevertheless, probing the electronic rearrangements at the molecular level remains challenging. Here, we explore the capabilities of soft X-ray ultrafast spectroscopy, as applied to the characterization of electronic dynamics during photoinduced intramolecular proton transfer in 2-(2'-hydroxyphenyl)-benzothiazole (HBT), shown in Fig. 1, as described by simulations of quantum dynamics including a full quantum treatment of all nuclear degrees of freedom.

At the molecular level, proton transfer involves electronic density changes directly at the proton donor and acceptor sites and beyond the donating and accepting groups, including polarization of the molecule and surrounding solvent environment. The effect of the proton on the local electronic structure of functional groups involved in proton exchange — occurring on the femtosecond timescale before, during, and after the elementary proton translocation — is still rather unexplored from an experimental point of view. *Ab initio* molecular dynamics simulations have provided insights into proton transport in bulk solvents through the von Grotthuss mechanism,[5] and acid dissociation dynamics.[6] In general, proton and electron transfer are thermodynamically coupled, yet kinetics might determine the overall relaxation and timescales of electron and proton rearrangements. So, fundamental questions remain to be addressed, including: Is a proton or a hydrogen atom transferred after photoexcitation? Is proton transfer driven by photoinduced electronic rearrangements? What electronic degrees of freedom are most critical for proton transfer? Those of the donating or accepting groups? Moreover, what nuclear motions are critical for hydrogen bonding,[7] and therefore essential for charge transfer events? Here, we address those fundamental questions in the study of electronic rearrangements due to enol-keto tautomerization in HBT after $S_0 \rightarrow S_1$ photoexcitation, which are dominated by the highest occupied to lowest occupied molecular orbital (HOMO → LUMO) transition that triggers intramolecular proton transfer along the intramolecular O-H···N hydrogen bond (Figure 1).

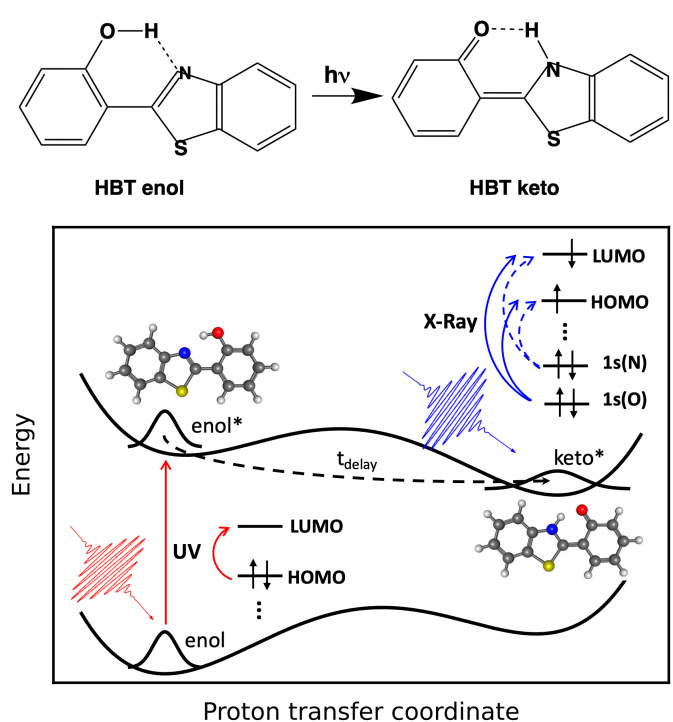

**Figure 1.** Schematic representation of pump-probe UV/X-ray spectroscopy to probe the HBT enol-keto tautomerization following photoexcitation (top). UV photoexcitation of the HOMO → LUMO transition initiates an ultrafast intramolecular proton transfer along the O-H···N hydrogen bond. A delayed soft K-edge X-ray pulse induces O/N 1s core → HOMO/LUMO transitions that probe transient electronic features of HBT.

The photoinduced enol-keto conversion of HBT occurs on the ultrafast timescale, as monitored by UV-vis[8] and UV-IR pump-probe spectroscopy.[9] Ultrafast spectroscopy has demonstrated that the molecular vibrations of HBT change in character when the C-O-H···N chemical bonds of the HBT-enol* state convert into the C=O···H-N configuration of the HBT-keto* state (* indicates photoexcited state).[9a-c, 9e] Furthermore, it was shown that low-frequency hydrogen-bonds are actively involved in the excited-state proton-transfer reaction.[8b, 10] In fact, a particular form of proton-coupled electron transfer (PCET) has been suggested to be an appropriate description since electronic structural rearrangements are concomitant with the proton transfer event.[1a, 1e, 9e, 11]

A combined study including polarization-resolved UV/IR pump-probe spectroscopy and time-dependent density functional theory (TDDFT) has shown that only a small amount of net positive charge is transferred to the benzothiazole ring when cis-enol* converts into cis-keto* tautomer.[9e] In fact, the benzothiazole side of HBT acquires a charge difference of about $0.34e$ (i.e., only $34\%$ of a full proton transfer) upon excited-state enol-keto tautomerization, which suggests significant electronic rearrangements through the molecular framework that mostly neutralize the proton translocation. However, direct spectroscopic evidence of the underlying electronic rearrangements remains challenging.

Soft X-ray spectroscopy provides unprecedented capabilities for probing the electronic structure of molecules undergoing photoinduced transformations with ultrafast and atomic (sub-molecular) resolution.[12] In recent years, major advances in steady-state and time-resolved soft-X-ray spectroscopy,[13] include the development of liquid jet technologies for sample delivery[14] at dedicated end stations of large scale facilities based on storage rings[15] and linear accelerators.[16] In addition, laser-based systems using extreme high-order harmonic generation of soft-X-ray pulses have become available.[17] Here, we analyze simulations of time-resolved X-ray Absorption Spectroscopy (TRXAS) to assess whether ultrafast soft-X-ray spectroscopy could provide a characterization of intramolecular PCET reactions.

Time-resolved XAS is particularly suitable for exploring PCET in HBT since it induces transitions of core electrons from 1s atomic orbitals localized on specific atoms involved in proton transfer to low-lying molecular orbitals (i.e., HOMO and LUMO) that are largely unaffected by PCET. Therefore, X-ray absorption provides access to changes in electronic states localized in proton donor and acceptor groups, which allows for ultrafast pump-probe schemes that can probe the evolution of the chemical reaction with atomic resolution.[18]

We focus on X-ray absorption spectroscopy using the K-edge absorption bands of nitrogen (N) and oxygen (O) to probe the transient electronic structure of HBT during the photoinduced enol-keto tautomerization induced by UV $S_0 \to S_1$ excitation. We follow the spectral signatures of O, while probing changes of its electronic environment, and N as it is protonated during formation of the keto product. We analyze the 1s → HOMO/LUMO transitions that can be probed by ultrafast X-ray absorption spectroscopy We follow the detailed evolution of the electronic excited state coupled to nuclear dynamics, including bond-breaking and bond-forming events described by a full quantum treatment of all nuclear degrees of freedom.

Monitoring changes in the K-edge bands of O and N could capture the evolution of the electronic structure at the proton donor and acceptor sites. A rigorous interpretation of the nuclear and electronic rearrangements responsible for changes in the XAS spectra can be provided by simulations of quantum dynamics as described by the tensor-train split operator Fourier transform (TT-SOFT) method.[19] TT-SOFT is a rigorous method for simulations of quantum dynamics that exploits matrix product state representations analogous to those employed by recently developed methods to simulate vibrational and fluorescence spectroscopy[20] and other methods for simulations of quantum dynamics and global optimization.[21] We have made the codes available in public domain.[23] Here, we demonstrate for the first time the capabilities of TT-SOFT as applied to simulations of UV/XAS pump-probe spectroscopy. We discuss how the UV and soft-X-ray pulse characteristics determine the outcome of an experiment, as could be implemented in state-of-the-art facilities that have become available in recent years.

## Results and Discussion

Figure 2 shows the evolution of the proton transfer coordinate $Q_{PT}$ associated with the enol-keto tautomerization, as it converts OH···N to O···HN. HBT starts in the enol* form, where the expectation value of the proton transfer coordinate is negative, and converts into the keto* form ($Q_{PT} \geq 0$) within $200$ fs, in agreement with previous results suggesting ESIPT.[9e, 23] The expectation value of the proton transfer coordinate remains in the range of values corresponding to the keto* isomer for the remainder of the examined dynamics, consistent with previous findings suggesting equilibration within $400$ to $600$ fs after photoexcitation of the system.[23b] Figure 2 also shows the time-dependent expectation value of the CCC bending mode, with an oscillation period of about $250$ fs comparable to the timescale for proton transfer. The time-dependent population of the keto isomer is also consistent with the ultrafast timescale for keto* formation within 50 fs after photoexcitation of the system (see Supporting Information).

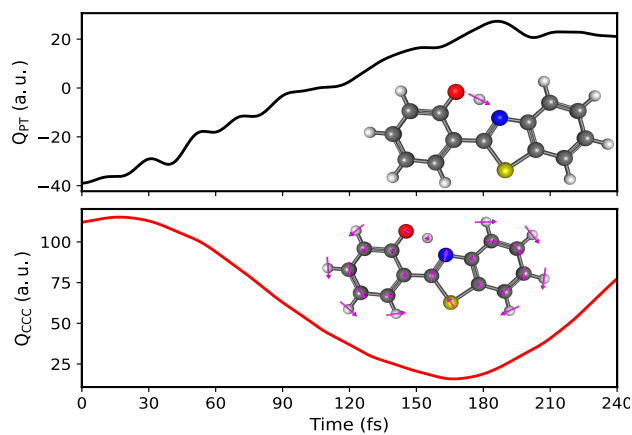

**Figure 2.** Dynamical evolution of the proton transfer (top panel) and CCC internal in-plane bending (bottom panel) large-amplitude modes after photoexcitation of HBT. Insets shows the normal mode displacements associated with each coordinate.

The reported quantum dynamics simulations are in good agreement with earlier studies, which supports the accuracy of our TT-SOFT simulations. UV/vis pump-probe measurements have demonstrated that the enol* → keto* reaction dynamics of HBT is strongly governed by low-frequency Raman-active modes that are impulsively excited by the electronic enol → enol* transition.[8b, 8c, 10, 23a, 24] In particular, the in-plane 255 cm$^{-1}$ mode plays a key role, as it modulates the O-H···N hydrogen bond distance and – thus – the reaction coordinate along which the H atom is transferred (see Fig. 2). The observed timescale for the enol* → keto* reaction dynamics has been ascribed to nuclear motions that in particular comprise this low-frequency mode, which compels photoexcited HBT to proceed on the excited state potential energy surface with a major excursion along this nuclear coordinate. A further refinement to this picture can be made, given additional pump-probe signal modulations have been ascribed to low-frequency modes at 113 cm$^{-1}$, 289 cm$^{-1}$ and 528 cm$^{-1}$. As our TT-SOFT method estimates the propagation dynamics of HBT in a full quantum treatment, our results can show to what extent these low-frequency Raman-active modes play a key role in the wavepacket dynamics in the enol* → keto* reaction dynamics and how these modes influence the femtosecond UV/XAS spectroscopic observables during the chemical reaction.

Figure 3 compares the simulated steady-state X-ray absorption near edge structure (XANES) spectra of HBT in the $S_0$ and $S_1$ electronic states, which correspond to the K-edge of nitrogen (left panel) and oxygen (right panel), respectively. The $S_0$ spectrum of N is characterized by an intense pre-edge peak centered at 387.9 eV and a broad band at ∼ 390.6 eV. Analysis of the natural transition orbitals (NTOs) shows the lower-energy peak corresponds mainly to the 1s(N)→LUMO transition (with small contributions from higher $\pi^*$ orbitals), whereas the higher energy band is dominated by the 1s core excitation to a delocalized $\pi^*$ orbital centered on the thiazole-ring. The O XANES spectrum, shown in Fig. 3 (right panel), is characterized by two main peaks located at 520.2 eV and 521.7 eV that arise from 1s core excitations from the 1s(O) orbital to the LUMO and a delocalized $\pi^*$ orbital centered on the hydroxyphenyl ring, respectively. The lowest energy transitions in both N and O spectra correspond to transitions to the same final state (*i.e.*, LUMO), whereas higher energy transitions reach excited states that are localized on different rings as determined by differences in cross sections of the 1s core excitations to the respective final states. We note that contributions from double excitations are neglected by the single-reference MOM/TDDFT method and are expected to contribute to the high-energy region of the spectra so they would require more advanced electronic structure calculation methods.

Changes in electronic character induced by the $S_0 \rightarrow S_1$ UV photoabsorption can be identified in the XANES spectra. Figure 3 shows the spectra corresponding to 1s core transitions after UV photoexcitation of the molecule to the $S_1$ excited state in the enol configuration (red lines). New pre-edge features dominated by the 1s core → HOMO transition are observed at 384.6 eV and 516.3 eV for N and oxygen K-edge, respectively. That well-resolved single peak feature arises from the vacancy of the HOMO created by the pump (see Fig. 1) and provides a distinct spectroscopic fingerprint of the electronic excited state. The spectral shifts between these pre-edge peaks correspond to the relative energy between 1s N and O orbitals (*vide infra*) since the final state is the same (*i.e.*, both transitions reach the same HOMO) for both the N and O pre-edge features. Rearrangements of the electronic density induced by the $S_0 \rightarrow S_1$ photoexcitation also change the peak positions and intensities of the remaining XANES bands. Note that the N 1s core → LUMO transition is red-shifted by 0.4 eV, whereas a ∼ 1 eV blue-shift is observed for the O atom 1s core → LUMO transition. These results suggest that the 1s core → HOMO/LUMO transitions of N and O provide valuable probes of the electronic excited state of the enol* tautomer.

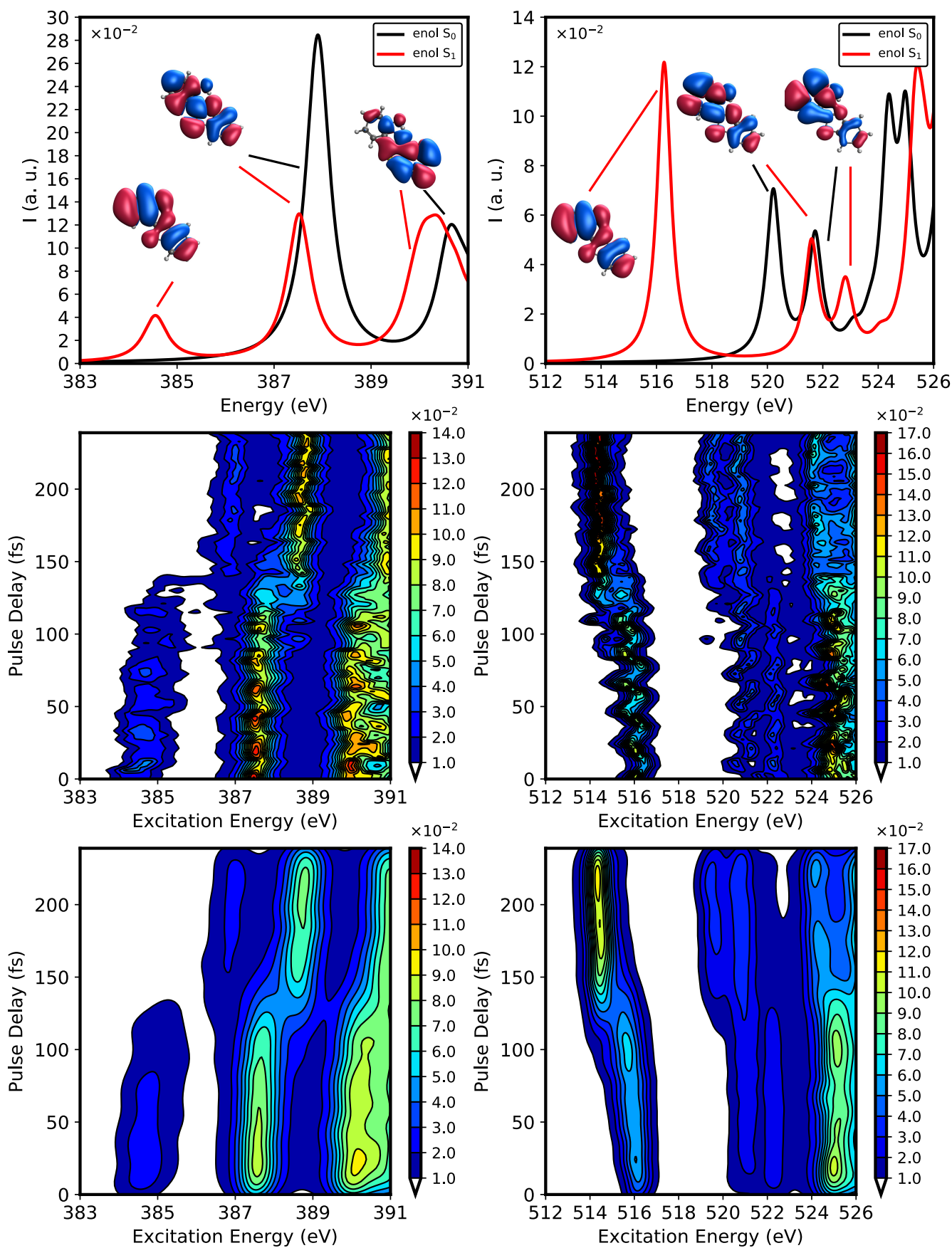

**Figure 3.** Top panels: K-edge nitrogen (left panel) and oxygen (right panel) XANES spectra for HBT in the enol tautomer in the $S_0$ ground state (black line) and following photoexcitation to the Franck-Condon region of enol* $S_1$ excited state (red line). Stick spectra were convoluted with Lorentzian functions with 500 meV FWHM. Inset: dominant NTOs of the 1s core excitation (isovalue: 0.02). Middle panels: Nitrogen K-edge (left) and oxygen K-edge (right) TRXAS following photoexcitation. Bottom panels: Instrumentally convoluted nitrogen K-edge (left) and oxygen K-edge (right) (normalized) TRXAS. The effective temporal and spectral resolution FWHM correspond to 20 fs and 0.5 eV, respectively.

Figure 3 shows transient X-ray absorption spectra computed after photoexcitation of the enol HBT tautomer, which provides evidence of the ability of TRXAS to probe the electronic structural changes during the photoinduced proton transfer dynamics. The nitrogen K-edge TRXAS (Fig. 3, left) provides clear fingerprints of nuclear dynamics in the $S_1$ excited state. A significant blue-shift (1–2 eV) is observed for the 1s core → HOMO/LUMO transitions of HBT in the enol* conformation when comparing the 0 fs delay time to the keto* conformation at 200 fs and beyond. The disappearance of peaks at 384.5 and 387.5 eV and the concomitant appearance of the peaks at 387 and 389 eV provides spectral signatures of the enol* → keto* transformation, predicted to occur in 100−150 fs, in excellent agreement with theory and earlier experiments.[9e, 23] Note that the timescale of the photoisomerization coincides with the minimum amplitude of the CCC bending mode (Fig. 2), which is consistent with the key role of the low-frequency reaction coordinate in regulation of the proton-transfer dynamics. The TRXAS spectra also show that both the peak positions and intensities oscillate with a period of about 20–24 fs, which suggests a tuning mode with a frequency of 1400−1700 cm$^{-1}$ (*i.e.*, within the expected frequency range of the OH/NH bending mode). Resonance Raman spectra of HBT show that modes involved in aromatic ring deformation and the O-H bending mode overlap in that spectral range, with intensities corresponding to pronounced displacements along those modes induced by the electronic excitation.[24] So, both of those modes are excited by the UV pump pulse and modulate the transient absorption spectrum, as shown in Fig. 3, which provides a detailed characterization of the photoisomerization process by UV/XAS pump-probe spectroscopy. Similar features are observed in the oxygen K-edge TRXAS spectrum (Fig. 3, right). Both the 1s core → HOMO and 1s core → LUMO transition frequencies exhibit a 1−1.5 eV shift on the timescale of 100–150 fs, which provides spectroscopic evidence of the ultrafast isomerization process with sub-ps resolution. These observations provide key insights into the dynamics of electronic structural rearrangements during the enol* → keto* tautomerization process.

Figure 4 shows that the frontier orbitals remain largely unaffected by the enol-keto tautomerization. Furthermore, Fig. 4 compares the time evolution of the HOMO and LUMO energies to the energies of 1s atomic orbitals of N and O during the isomerization dynamics. The time-dependent orbital energies provide a fundamental understanding of the dynamical features of TRXAS probing the enol*→ keto* isomerization. Spectral shifts for those transitions report directly on changes in the frontier orbital energies relative to the 1s core atomic orbital energies since the peaks correspond to 1s core → HOMO/LUMO transitions, within a frozen-core approximation. Remarkably, the 1s core orbitals exhibit a significant energy shift, whereas changes in the frontier orbital energies are much smaller. Therefore, the resulting shifts report on localized changes in the electronic structure of the chromophore with sub-molecular (atomic) spatial resolution.

We find that the enol-keto tautomerization stabilizes the N 1s orbital and destabilizes the O 1s core level, in excellent correlation with the observed changes in the N and O K-edge TRXAS peaks, which become red and blue shifted, respectively (Fig. 3). In addition, Fig. 4 shows that changes in the energy levels of the HOMO and LUMO frontier orbitals, which play a key role in the excited state ultrafast tautomerization, are rather subtle. Similar findings have been reported for 2-(2'-hydroxyphenyl)-benzotriazole[25] and methylsalicylate.[26]

The time-dependent Mulliken charges, displayed on Fig. 4, indicate that changes of the core orbital energies are correlated with the rearrangement of electronic charges induced by the enol* → keto* tautomerization. The O negative charge increases (in magnitude) during the tautomerization process, whereas the N atom becomes more neutral. The increase in electronic density around the O atom destabilizes the 1s orbital, which leads to a smaller energy gap between the 1s core orbital and the delocalized frontier orbitals. The opposite is observed for the N atom. As such, the blue and red shifts of the N and O K-edge TRXAS peaks, shown in Fig. 3, can be traced back to changes in the electronic charge distribution induced by the tautomerization reaction.

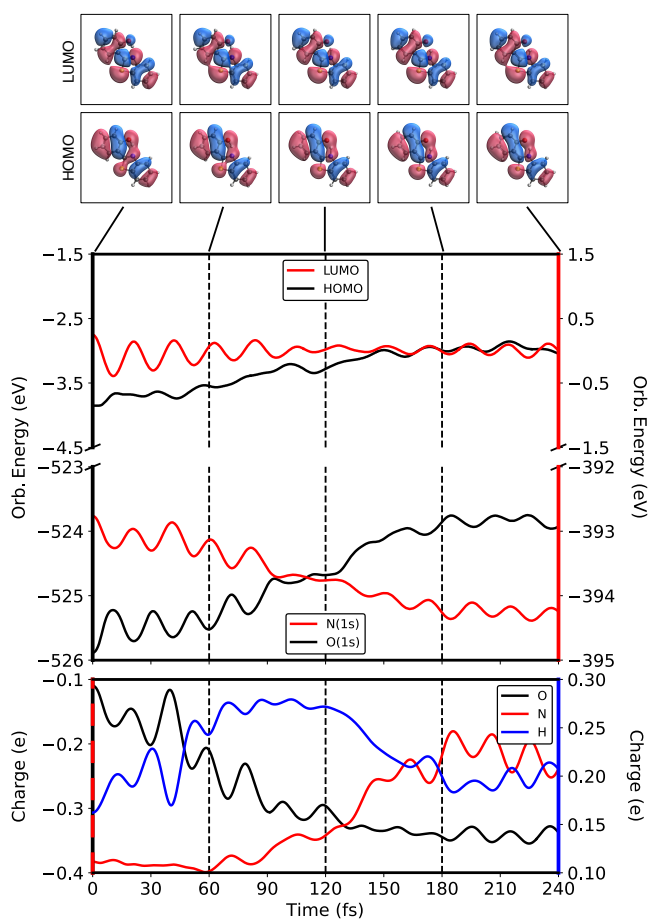

**Figure 4.** Top panel: Time evolution of the HOMO/LUMO and core 1s orbital energy of O and N atoms. Bottom panel: Time evolution of the Mulliken charges on the O, N, and transferring H atoms. Note the different scale of the data (as indicated by the colored border of the axis). Insets at the top show the HOMO and LUMO frontier orbitals, which indicate subtle changes during the enol* → keto* tautomerization reaction (isovalue = 0.02 Å$^{-3}$).

Figure 4 shows transient changes in the charge on the transferring H as the OH bond is broken and the NH bond is formed. H becomes most positive halfway through the transfer process (at about 120 fs) before the N-H bond is formed. That intermediate stage exhibits a strong attenuation of the pre-edge spectral peak. The N 1s core→HOMO TRXAS signal recovers only after the N-H bond is formed, which demonstrates the sensitivity of the TRXAS peak intensities to electronic reorganization dynamics, in addition to the previously discussed energy shifts. Therefore, we find that the evolution of the TRXAS spectra is able to probe the detailed dynamics of intramolecular proton-coupled electron transfer,[9e] including changes in the electronic environment of the proton donor and proton acceptor functional groups within 200 fs. We remark that the total charge on the benzothiazole and hydroxyphenyl ring along the isomerization process follows similar trends to the charge on O and N, respectively (see Supporting Information).

Figures 3 and 4 show the sensitivity of the nitrogen and oxygen K-edge XAS bands to the dynamics of intramolecular electronic rearrangements due to the tautomerization of HBT, assuming infinite resolution. In practice, we anticipate spectrometer devices might offer an effective temporal resolution of 20 fs (*i.e.*, spectral resolution of 0.5 eV), as shown in Fig. 3. Clearly, with that limited resolution, the fastest nuclear coherences of fingerprint vibrational modes would be averaged out. Nevertheless, the essential, dominant contribution of the hydrogen-bond modulation mode that governs the timescale of the enol*→keto* conversion at about 120 fs would remain clearly discernible. Hence, we anticipate that experiments that probe transient electronic structural dynamics of PCET can provide unique insights beyond the capabilities of conventional ultrafast spectroscopic methods.

## Conclusion

We have investigated the ability of time-resolved X-ray absorption spectroscopy at both the nitrogen and oxygen K-edge to resolve the ultrafast enol-to-keto isomerization photophysics in HBT upon photoexcitation. We find that TRXAS can resolve the longstanding question of whether the photoinduced HBT tautomerization proceeds via excited state PCET or intramolecular hydrogen transfer. The reported simulations show that UV-pump soft-X-ray-probe spectroscopy can probe the nitrogen and oxygen K-edge bands to provide a detailed characterization of the ultrafast dynamics of electronic structural rearrangements during the photoinduced enol*→keto* tautomerization of HBT. The reported results show that the main driving force of the reaction results from electronic structural rearrangements initiated by UV photoexcitation. The electronic excitation triggers relaxation of nuclear coordinates through a sequence of stages, starting with displacement of the CCC internal bending bond and bringing together the proton donor (OH) and proton acceptor (N) functional groups. The resulting displacement polarizes the O-H bond by strengthening the OH⋯N hydrogen-bond, which increases the positive atomic charge of H and the negative atomic charge of O and partially reduces the negative atomic charge of N by delocalization through conjugated double-bonds within 120 fs of HBT photoexcitation (Fig. 4). The proton is then transferred and forms a covalent bond with N. The resulting changes in electronic density change the bond order of proton donor/acceptor groups and evolve the hydroxyphenyl and benzothiazole rings into resonance structures corresponding to the keto isomer.

We have shown that the TT-SOFT dynamics simulates the nuclear motion, including quantum delocalization of the transferring proton through the low-barrier hydrogen-bond of the $S_1$ state, while the MOM/TDDFT method resolves the electronic transitions of TRXAS. Therefore, the combination of TT-SOFT and MOM/TDDFT methods has allowed us to track both electronic and nuclear rearrangements fully quantum mechanically and resolve the PCET nature of the reaction mechanism. The methodology is quite general and applicable to a wide-range of molecules beyond HBT. Chemical systems amenable to TT-SOFT simulations are widespread, including other reactive coordinate systems with quantum baths and hydrogen-bonding in DNA base pairs, water molecules, and Zundel cations. Therefore, we anticipate that TT-SOFT/MOM simulations of pump-probe spectra can provide valuable interpretations of UV pump-X-ray probe spectra of chemical processes where analogous dynamics of electronic and nuclear rearrangements determine the timescale and nature of reaction mechanisms as well as the outcome of photoinduced reactions.

PCET for HBT is not fully concerted, but rather sequential, with charge migrations occurring on distinctly different times for the oxygen donating atom, for the inner proton and the nitrogen accepting atom. Initially, during the first 120 femtoseconds of the PCET process, when the O⋯N distance is shortened due to the internal bending motion between the hydroxyphenyl and benzothiazole units and the H atom has transferred halfway, electronegative charge density at the oxygen atom increases, while that of the inner H atom decreases (making it more positive). In the next 120 femtoseconds of the PCET process, the O⋯N distance is lengthened again, and the H atom electronegative density increases, while that of the nitrogen atom decreases. These results show the interplay between charge migration, hydrogen bond modulation motions, and proton motion. These results are expected to be of high relevance for other PCET processes, and ultrafast soft-X-ray spectroscopy will be a highly valuable tool to discern the different steps in PCET.

## Acknowledgements

M. B. S. acknowledges financial support from the Yale Quantum Institute Postdoctoral Fellowship. V. S. B. acknowledges support from the NSF Grant no. CHE-1900160 and high-performance computing time from NERSC. E. T. J. Nibbering acknowledges support from the European Research Council (ERC) under the European Union's Horizon 2020 research and innovation programme (ERC Grant Agreement N° 788704; E.T.J.N.). This work is dedicated to Thomas Elsaesser for his contributions to ultrafast science in general and to ultrafast photoinduced intramolecular proton transfer processes in particular for a timeline of more than four decades.

**Keywords:** proton coupled electron transfer • femtosecond soft-X-ray spectroscopy • pump probe spectroscopy • tensor networks • quantum dynamics

**Entry for the Table of Contents**

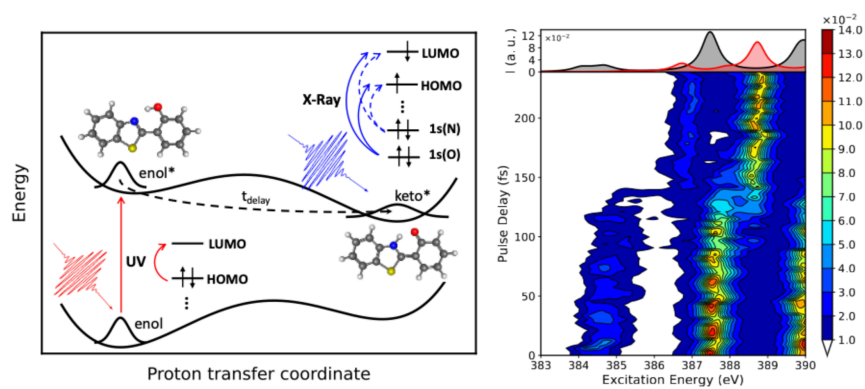

Proton-coupled electron transfer (PCET) is essential to a wide range of chemical processes, from photosynthesis to hydrogen fuel cells. We present a fully quantum simulation study of photoinduced enol* → keto* tautomerization that demonstrates femtosecond UV/X-ray pump-probe spectroscopy provides an unprecedented view into the charge migration and ultrafast electronic structure rearrangements that play a key role in PCET.

Institute and/or researcher Twitter usernames: @MichelineSoley, @PabloEVidela, @VictorB59730354

# Supporting Information - Ultrafast Charge Migration Dynamics in Enol Keto Tautomerization Monitored with a Local Soft-X-Ray Probe

Micheline B. Soley,[+,*,[a,b]] Pablo E. Videla,[+,*,[a,c]] Erik T. J. Nibbering,[d] and Victor S. Batista*[a-c]

[a]  Dr. M. B. Soley,[+] Dr. P. E. Videla,[+] Prof. V. S. Batista*
     Department of Chemistry, Yale University
     P.O. Box 208107, New Haven, CT 06520-8107 (USA)
     E-mail: micheline.soley@yale.edu
             pablo.videla@yale.edu
             victor.batista@yale.edu
[b]  Dr. M. B. Soley, Prof. V. S. Batista
     Yale Quantum Institute, Yale University
     P.O. Box 27394, West Haven, CT 06520-8263 (USA)
[c]  Dr. P. E. Videla, Prof. V. S. Batista
     Energy Sciences Institute, Yale University
     P. O. Box 27394, West Haven, CT 06516-7394 (USA)
[d]  Dr. E. T. J. Nibbering
     Max Born Institute for Nonlinear Optics and
     Short Pulse Spectroscopy
     Max Born Strasse 2A, 12489 Berlin (Germany)
[+]  These authors contributed equally to this work.

## Methods

In order to provide a highly accurate probe of the HBT isomerization mechanism, we simulate the UV-pump/soft X-ray probe spectroscopy with quantum treatment of all all degrees of freedom as follows: The *ab initio* ground and excited state potential energy surfaces are generated at the TDDFT level of theory, and the ground state wavefunction is identified according to imaginary time propagation in the ground electronic state and instantaneously shifted to the excited electronic state to model the UV pump, with quantum dynamics of the molecule computed according to the low-rank tensor train Split Operator Fourier Transform (TT-SOFT) method. The resulting time-resolved X-ray spectra are computed as a convolution of the wavepacket with K-edge core excitations with transition energies and dipole strengths determined with the Maximum Overlap Method (MOM) and Time-Dependent DFT (TDDFT).

## Potential Energy Surface

We describe the dynamics of HBT following photoexcitation in terms of full-dimensional ground and excited state potentials based on a reaction surface approach,[1] similar to the ones implemented in earlier studies of ESIPT.[2] In the present work, we use mass-weighted normal modes, determined at the transition state of the $S_1$ electronic state, as generalized coordinates to characterized the proton transfer dynamics. The normal mode displacement associated with the imaginary frequency is selected as the proton transfer reactive coordinate $Q_1$. Additionally, the normal mode with predominantly components along the CCC internal in-plane bending mode $Q_2$ is included. The latter modulates the N−O distance between the proton donor and acceptor moieties and is expected to be strongly coupled to $Q_1$. Figure S1 shows the normal modes displacement.

Each full-dimensional potential energy surface $V(Q_1, Q_2, \mathbf{z})$ is constructed as a quadratic expansion around a 2-dimensional reaction surface potential $V_0(Q_1, Q_2)$,

$$V(Q_1, Q_2, \mathbf{z}) = V_0(Q_1, Q_2) + \frac{1}{2}[\mathbf{z} - \mathbf{z}_0(Q_1, Q_2)] \cdot \mathbf{K}(Q_1, Q_2) \cdot [\mathbf{z} - \mathbf{z}_0(Q_1, Q_2)], \qquad (1)$$

where $\mathbf{z}$ represents the other 67 vibrational normal modes that are described as locally harmonic oscillators with force constants $\mathbf{K}(Q_1, Q_2)$ and equilibrium positions $\mathbf{z}_0(Q_1, Q_2)$ parametrized by the reaction coordinates $Q_{1,2}$. The *ab initio* ground and excited state surfaces $V_0(Q_1, Q_2)$ are determined by fully optimizing the geometry of the system with respect to all other degrees of freedom $\mathbf{z}$, subject to the constraint of a fixed value of $Q_{1,2}$, at 12 and 20 equally spaced points in the range $Q_1 \in [-50.0, 60.0]$ a.u. and $Q_2 \in [-140.0, 240.0]$ a.u., respectively. The equilibrium coordinates $\mathbf{z}_0(Q_1, Q_2)$ represent the normal mode displacements of the bath modes relative to the reference configuration and are computed by projecting the optimized $Q_{1,2}$-constrained geometry onto the normal modes displacement vectors determined at the transition state of the $S_1$ electronic state. The force constants $\mathbf{K}(Q_1, Q_2)$ representing the bath couplings are determined by computing the (mass-weighted) Hessian at each optimized $Q_{1,2}$-constrained geometry and projecting out the contribution due to the $Q_{1,2}$ reactive coordinates.[1b]

A hierarchical set of approximations can be applied to Eq. (1). If one neglects the coupling between the different bath modes (i.e. neglecting non-diagonal elements in $\mathbf{K}(Q_1, Q_2)$), the potential reduces to

$$V(Q_1, Q_2, \mathbf{z}) = V_0(Q_1, Q_2) + \sum_j \frac{1}{2} K_{jj}(Q_1, Q_2)[z_j - z_{j,0}(Q_1, Q_2)]^2. \quad (2)$$

If one further neglects the dependence of $\mathbf{K}(Q_1, Q_2)$ on $Q_{1,2}$, the potential reduces to

$$V(Q_1, Q_2, \mathbf{z}) = V_0(Q_1, Q_2) + \sum_j \frac{1}{2} \omega_j^2 [z_j - z_{j,0}(Q_1, Q_2)]^2 \quad (3)$$

with $\omega_j^2 = K_{jj}(Q_1 = 0, Q_2 = 0)$. If one also neglects the $Q_{1,2}$-dependence on $\mathbf{z}_0(Q_1, Q_2)$, the potential reduces to

$$V(Q_1, Q_2, \mathbf{z}) = V_0(Q_1, Q_2) + \sum_j \frac{1}{2} \omega_j^2 z_j^2 \quad (4)$$

where we have used the fact that $z_{j,0}(Q_1 = 0, Q_2 = 0) = 0$.

All excited-state calculations were performed at the TDDFT level of theory using the $\omega$B97XD functional[3] and cc-pvdz basis set,[4] as implemented in Gaussian 16.[5] Solvent effects (dichloromethane) were included implicitly through the PCM polarizable continuum model.[6]

These coarsely grained TDDFT potentials are used to generate a finer position-space grid to improve the accuracy of the quantum dynamics simulation in the reactive potential $V(Q_1, Q_2)$. Potential values at positions intermediate between grid points are determined according to cubic splines, and potential values at positions outside of the original position space domain are defined to be equal to the value at the edge of the original domain. The resulting reactive potential energy surface is shown in Fig. S1 with representative large-amplitude mode dependent displacements and coupling constants in Fig. S2.

To reduce the computational cost associated with evaluation of the potential at all points on the 69-dimensional position space grid, the potential is compressed according to the low-rank tensor train approximation,[7] a form of matrix product states[8] defined as follows:

$$Q(i_1, \ldots, i_d) \approx \sum_{\alpha_1=1}^{r_1} \sum_{\alpha_2=1}^{r_2} \ldots \sum_{\alpha_{d-1}=1}^{r_{d-1}} A_1(1, i_1, \alpha_1) A_2(\alpha_1, i_2, \alpha_2) \ldots A_d(\alpha_{d-1}, i_d, 1), \quad (5)$$

where $i_j$ corresponds to the position space coordinates in physical dimension $j$ and $A_j$ is a three-mode tensor contracted to the neighboring tensors by indices $\alpha_{j-1}$ and $\alpha_j$. The low-rank tensor train decomposition is adaptively computed via the cross approximation.[1a]

## Dynamics

To monitor atomic motion in the HBT isomerization process, the *ab initio* potential energy surfaces are used to simulate dynamics of the HBT molecule with tensor-train Split Operator Fourier Transfrom (TT-SOFT) quantum dynamics. This method is employed to accurately incorporate quantum effects on the isomerization process, as it treats all of the molecule's 69 normal modes fully quantum mechanically.[9] Tensor train manipulations are performed with Oseledet's TT-Toolbox for fast interpolation procedure[10] via the cross approximation,[7a] which avoids calculation of the potential and wavepackets at all points on the position space grid. Tensor trains are evaluated with a relative accuracy parameter of $\epsilon = 10^{-14}$ and a maximum rank of $r_\text{max} = 10$ to yield an accurate, low-cost representation of physical quantities such as the potential energy surface, wavepacket, and propagators.

TT-SOFT is employed for imaginary time propagation in the ground electronic state potential energy surface $S_0$ to determine the initial ground state wavepacket. The ground state wavepacket is then instantaneously excited to the excited electronic state $S_1$, whereupon real time propagation is used to investigate excited state intramolecular proton transfer (ESIPT).

Both imaginary and real time propagation are carried out for $N_\tau = 800$ time steps of length $\tau = 12.5$ a.u. to accurately follow the motion of short wavelength normal modes, ensure convergence of the ground state wavefunction, and follow the isomerization process, which is known experimentally to occur within $T = 250$ fs.[2a, 2c, 11] The position space grid in each direction is discretized into $N_p = 2^5$ equally spaced grid points for sufficient resolution of the potential energy surface. The large-amplitude modes are considered in the domain $Q_1 \in \{-100, 100\}$ a.u. and $Q_2 \in \{-250, 250\}$ a.u. and the bath modes are considered in the domain $Q_i \in \left\{ -\frac{10}{\sqrt{\omega_i}}, \frac{10}{\sqrt{\omega_i}} \right\}$ to include the full extent of the wavepacket motion. Prior to imaginary time propagation, the wavepacket is initialized as a Gaussian wavefunction with a noise factor

$$\psi_0(\mathbf{x}) = \prod_{j=1}^{D} \sqrt[4]{\frac{m\omega_j}{\pi}} e^{\frac{m\omega_j}{2\sigma_j^2}(x_j - x_0)^2} + 10^{-16} \Theta(j - 50) \quad (6)$$

$$\Theta(j) = \begin{cases} 1 & j \geq 0 \\ 0 & j < 0 \end{cases} \quad (7)$$

for particle mass $m = 1$ and frequency $\omega_j$, width parameter $\sigma_j$, and center $x_j$ of normal mode $j$ for $D$ dimensions. To initialize the wavepacket close to the true ground state wavefunction while encouraging sampling of nonzero points in the cross approximation interpolation procedure, the Gaussian is initially centered at a position slightly shifted from the minimal position of the ground state potential energy surface to facilitate fast identification of the ground state wavefunction ($x_0 = -32.5$ a.u., $x_1 = 123.515625$ a.u.,

$x_i = z_{i,0}(0,0)$ for $i > 2$). The Gaussian widths of the large-amplitude modes are defined by the harmonic approximation for the ground state well ($\sigma_1 = 22.473$ a.u. and $\sigma_2 = 53.132$ a.u.) and those of the bath modes are defined in terms of the couplings ($\sigma_i = \sqrt{\dfrac{2}{\omega_i}}$ for $i > 2$).

The HBT motion represented by propagation of the initial wavepacket is evaluated by computing the average expectation value of position of each coordinate, which is given by the inner product of the wavefunction and the corresponding position-space operator acting on the wavefunction.

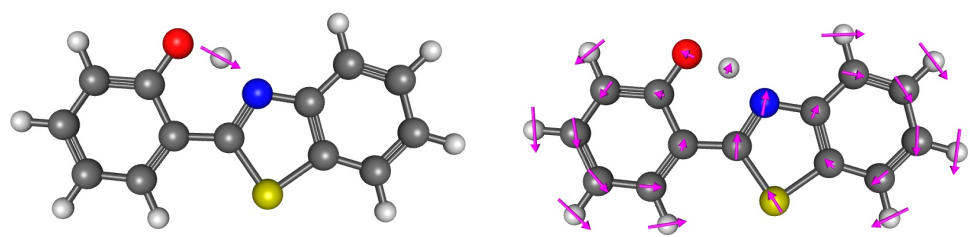

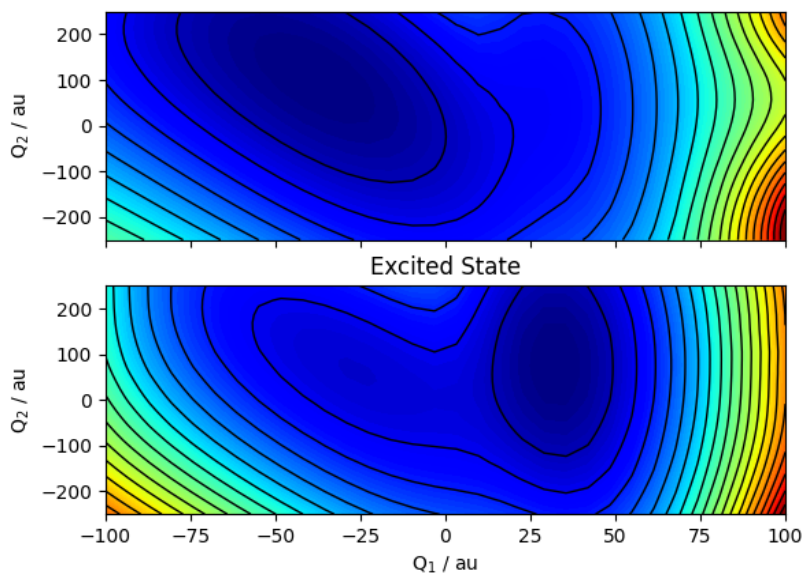

**Figure S1.** Top: Normal mode displacements of the large-amplitude modes associated with the proton transfer coordinate $Q_1$ and CCC internal in-plane bending coordinate $Q_2$. Bottom: Reactive potential energy surfaces $V_0(Q_1, Q_2)$ of the $S_0$ and $S_1$ electronic states as a function of the large-amplitude modes $Q_1$ and $Q_2$.

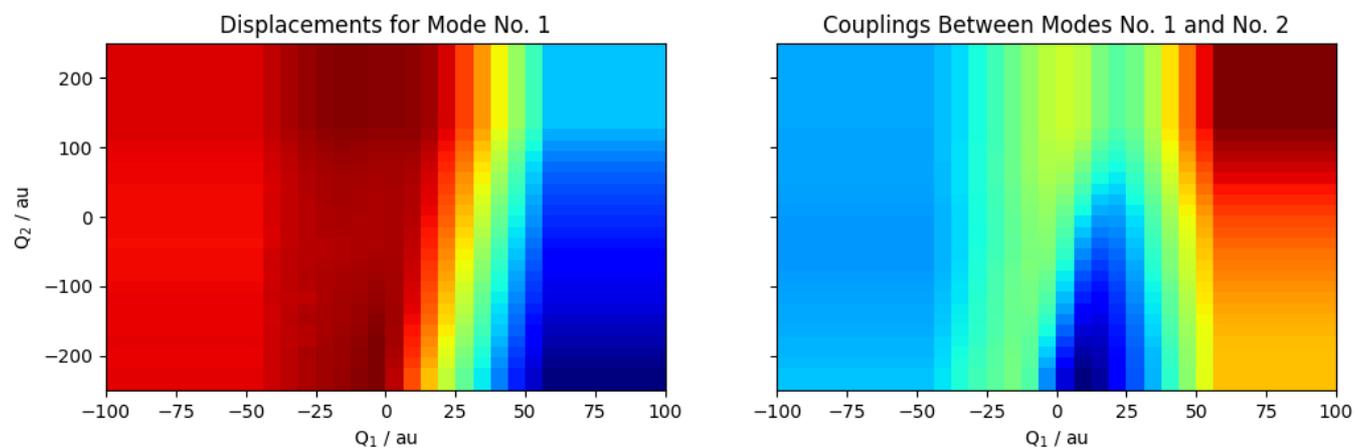

**Figure S2.** Representative ab initio $Q_1$ and $Q_2$ large-amplitude mode dependent displacement (left) and coupling constant (right). In the adiabatic approximation, coupling constants are fixed to their $(Q_1, Q_2) = (0,0)$ a.u. value.

## Time-Resolved UV-Pump/X-Ray Probe Spectra

The overall behavior of the HBT molecule during isomerization is investigated via prediction of the transient X-ray absorption spectrum (TRAXS). The transient X-ray spectral lineshape is modeled as an incoherent superposition of core excitations weighted by the magnitude of the nuclear wavepacket at a given time instant, namely

$$I(\omega, t) = \sum_k \int d\mathbf{Q} |\Psi(\mathbf{Q}, t)|^2 |\mu_k(\mathbf{Q})|^2 \delta(\omega - \omega_k(\mathbf{Q})) \quad (8)$$

where $\Psi(\mathbf{Q}, t)$ represents the wavepacket at time $t$ and nuclear coordinates $\mathbf{Q}$, $\omega$ is the photon energy, and $\omega_k(\mathbf{Q})$ and $\mu_k(\mathbf{Q})$ correspond to the $k$-th electronic transition energy and transition strength, respectively.

In the present work, for simplicity we approximate the nuclear density along the bath modes by a Dirac delta function at the expectation value of the position operator, given by $\bar{\mathbf{z}} = \int d\mathbf{Q} |\Psi(\mathbf{Q}, t)|^2 \mathbf{z}$. Therefore, the transient X-ray spectrum was calculated as

$$I(\omega, t) = \sum_k \int dQ_1 dQ_2 |\Psi(Q_1, Q_2, \bar{\mathbf{z}}, t)|^2 |\mu_k(Q_1, Q_2, \bar{\mathbf{z}})|^2 \delta(\omega - \omega_k(Q_1, Q_2, \bar{\mathbf{z}})) \quad (9)$$

The evaluation of the integral over the $Q_{1,2}$ quantum coordinates was performed by Monte Carlo sampling, using the ratio between the nuclear density as the acceptance parameter in the Metropolis algorithm.[12] Forty configurations were sufficient to converge the spectra. The spectral sticks were broadened with a Lorentzian envelope with full width at half maximum of $0.25$ eV to approximate the core-hole lifetime broadening of each atom and take into account experimental resolution.

The X-ray transition energies and dipole strengths for the excited $S_1$ state were obtained by a combined Maximum Overlap Method (MOM)[13] and Time-Dependent DFT[14] approach as implemented in the Q-Chem package.[15] The MOM strategy was used to obtained the $S_1$ electronic wavefunction, which was subsequently used as a reference state to performed TDDFT core-excitations. The initial guess for the MOM-SCF procedure was taken as $[\text{HOMO}(\alpha)]^1 [\text{LUMO}(\beta)]^1$ to resemble the hole and particle NTO of the transition from the ground state to the valence-excited state. We used the initial MOM (IMOM) method[13] instead of standard MOM to ensure convergence to the desired $S_1$ reference electronic state. From the $S_1$ reference state, TDDFT calculations were performed with a reduced single-excitation space that included the nitrogen or oxygen 1s orbital and all the virtual orbitals to obtain 30 core-valence X-ray excitations. All calculations were performed using the $\omega$B97XD functional[3] and cc-pvdz basis set,[4] with dichloromethane employed as implicit solvent through PCM.[6]

## Fully Quantum Treatment of 69D Dynamics

The fully quantum treatment of HBT dynamics successfully reproduced known behaviors of the HBT isomerization reaction. The average trajectory was found to be in agreement with previous experimental and theoretical findings, as shown in Fig. S3. The expectation value of the OH normal mode coordinate $Q_1$ shifted from a position in the enol* well to the keto* well within $200$ fs, in agreement with previous results that suggest excited state intramolecular proton transfer (ESIPT) occurs within $200$ fs.[2a, 2c, 11] The expectation value of the OH stretch coordinate remained in the keto* well for the remainder of the examined dynamics, which agrees with previous findings that suggest equilibrium is achieved within $400$ to $600$ fs. The expectation value of the large-amplitude mode coordinate corresponding to the C-C-C stretch exhibited an oscillation with a period on the order of $250$ fs, which suggests proton/hydrogen transfer is accompanied by a ring-bending motion on the same timescale. The theoretically determined population of keto isomers supported previous findings that significant keto* formation occurs within the first $50$ fs of dynamics and features oscillations on timescales below $50$ fs.[2a]

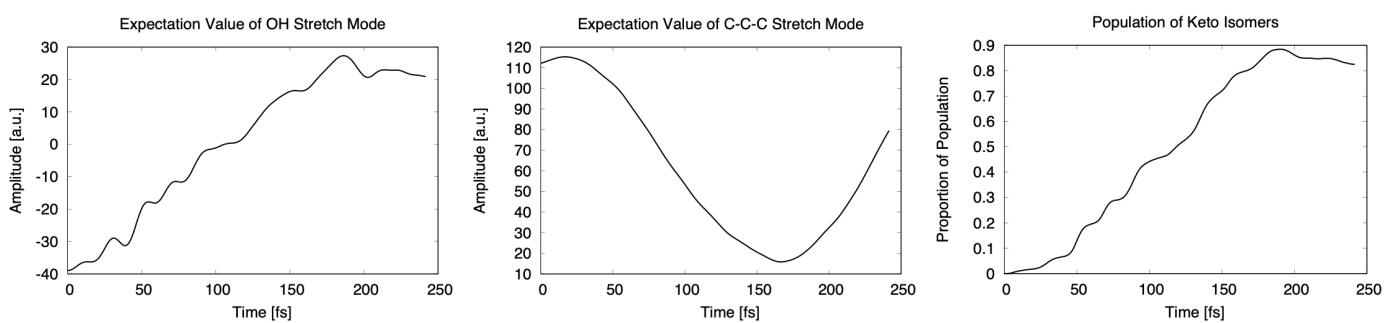

**Figure S3.** Expectation value of position of the large amplitude mode coordinates (left) $Q_1$ and (center) $Q_2$ and population of the keto isomer (right) confirmed the expected dynamics of HBT isomerization.

Further support for the accuracy of the quantum simulation was given by the probability density evaluated at the large amplitude mode coordinates (holding bath mode positions fixed at their expectation value, see Fig. S4). The tunneling of the wavepacket out of the enol* well in the first $50$ fs and into the keto* well within the first $250$ fs was in agreement with previous experimental and theoretical results, as was the equilibration of the wavepacket in the enol* well for the remainder of the simulated dynamics.

Furthermore, the probability density demonstrated tunneling plays a significant role in the isomerization process. At 90 fs − 240 fs after the instantaneous pump, both the keto* and enol* isomer states are observed, in line with the hypothesis that HBT isomerization does not occur in a concerted fashion and involves quantum effects in which both isomers participate in the isomerization process. The dynamics therefore reflected the full range of behaviors involved in HBT isomerization, including OH and CCC normal mode oscillations and tunneling effects.

## Evaluation of Potential and Spectrum Approximations

The accuracy of the 69-dimensional potential energy surface was evaluated through comparisons of dynamics and transient absorption X-ray spectra for varying rank, numbers of quantum bath modes, approximations to the full potential energy surface, and trajectory sampling.

The accuracy of the low-rank approximation to the reactive potential is verified in Fig. S5. Cross approximation of the reactive potential without subsequent rounding to a lower rank yields a reactive potential in agreement with the rank four result, which suggests the maximal rank 10 used to generate the dynamics is sufficient to accurately reproduce the potential energy surface. Similar results are obtained for the coupling and displacement parameters, as shown for the examples in Fig. S6.

Evaluation of the dynamics for differing numbers of quantum bath modes supports the need for inclusion of a large number of bath modes for successful simulation of HBT isomerization. As shown in Fig. S7, inclusion of only the reactive coordinates is insufficient to model irreversible formation of the keto isomer, as the molecular system is unable to dissipate energy into the bath coordinates. This results in periodic reformation of the enol isomer. As the number of bath modes included is increased, HBT approaches aperiodic formation of the keto isomer as expected from previous UV-IR experiments.

The need to include a large number of bath modes to simulate equilibration of the keto isomer is also supported by visualization of the probability density as a function of the number of large-amplitude modes, as shown in Figs. 4 and 8. In lower-dimensional examples, the wavepacket tunnels back and forth between the two isomeric wells, whereas the 69D result yields the expected irreversible keto formation.

In Figure S9, we present the time-resolved (TR) N and O K-edge XANES of HBT after photoexcitation for the 2D model, without bath modes. The TR spectra is characterized by well-separated peaks near $384$eV and $388$eV in the N K-edge spectrum and $516$eV, $520$eV, and $522$eV in the O K-edge spectrum. In the absence of the quantum bath modes, the peak oscillations closely agree with the oscillations of the expectation values of the large-amplitude mode coordinates on which the spectra are based. As in the expectation value of position of the coordinates, the spectral peaks at the initial and final times examined agree, which suggests a reformation of reactants indicative of the absence of the needed quantum bath.

Fig. S10 illustrate the improved accuracy of the TRXAS spectra from Monte Carlo sampling of the probability density in contrast to inclusion only of the average trajectory. Whereas the TRXAS computed via Monte Carlo sampling includes paired peaks at the reactant and product frequencies during the isomerization process in the O and N K-edge spectra, the TRXAS computed only with average trajectories yields a single peak that interpolates between these two values. This suggests sampling of the probability density beyond the average trajectory is required to demonstrate the full role of quantum effects in the isomerization process. Nonetheless, the average trajectory is valuable in isolating the reactant and product frequencies in the TRXAS as well as the overall time-dependent behavior of the molecule.

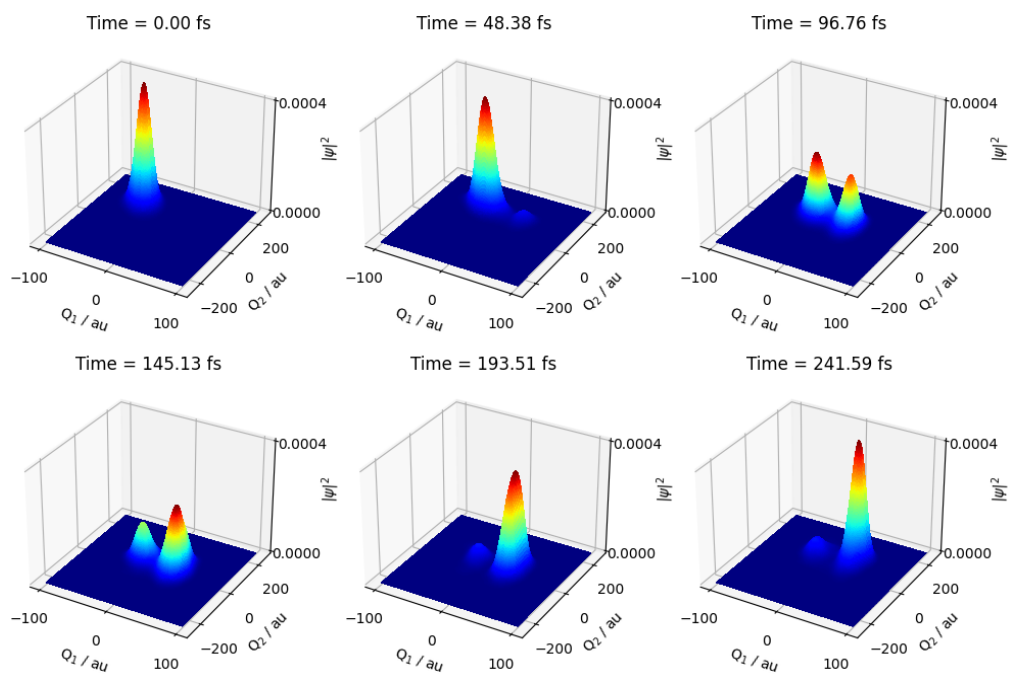

**Figure S4.** Probability density in terms of large amplitude mode coordinates $Q_1$ and $Q_2$ during the HBT isomerization process with integration over quantum bath modes.

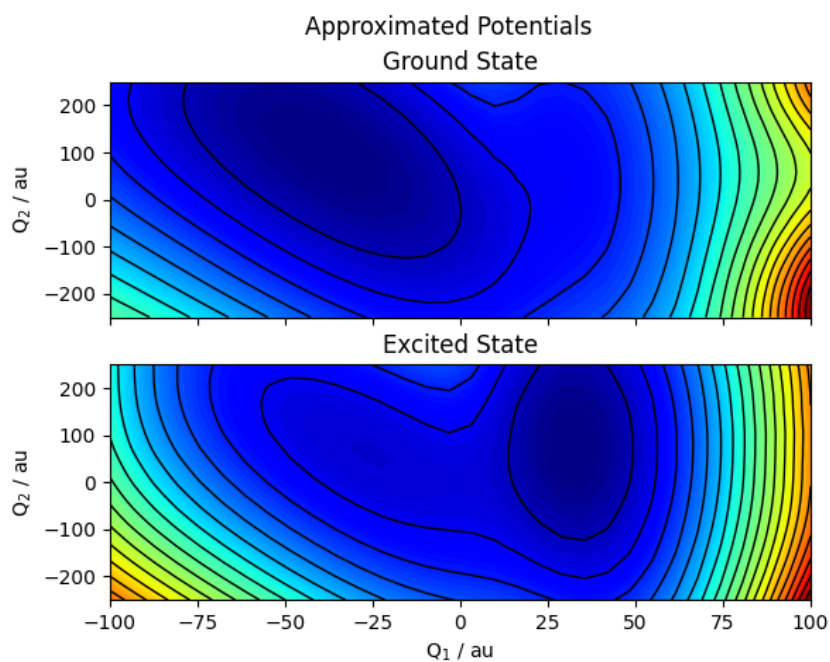

**Figure S5.** Low-rank tensor train approximation of reactive potential.

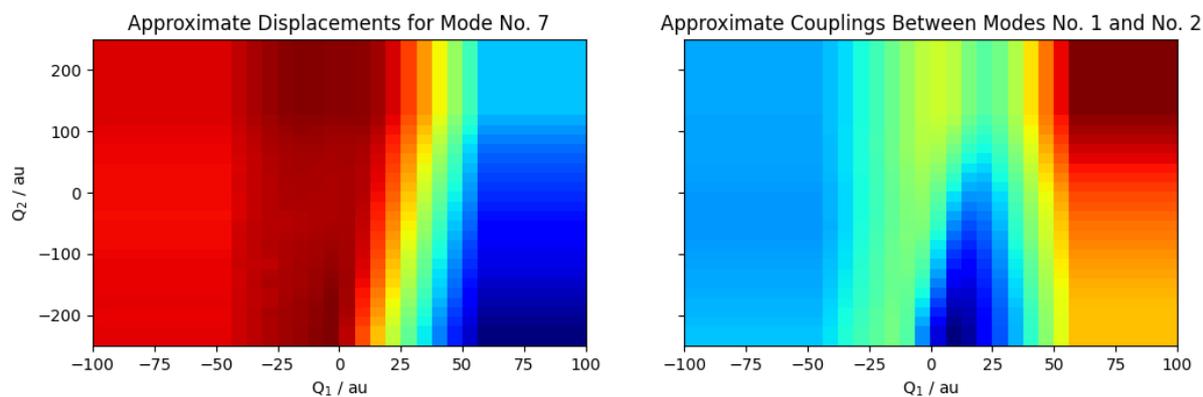

**Figure S6.** Low-rank tensor train approximation of an example large-amplitude mode dependent (left) displacement and (right) coupling constant.

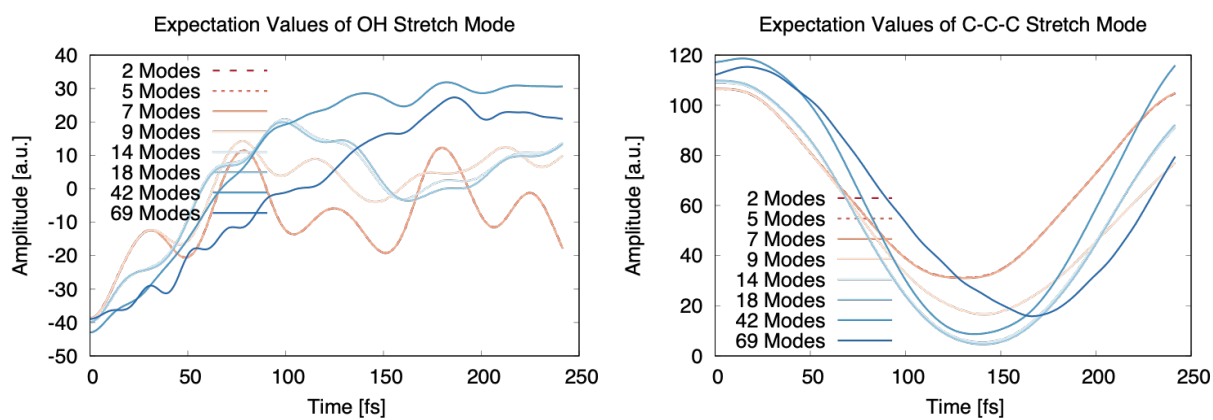

**Figure S7.** Expectation value of the large-amplitude mode coordinates $Q_1$ and $Q_2$ with inclusion of zero to 67 bath modes.

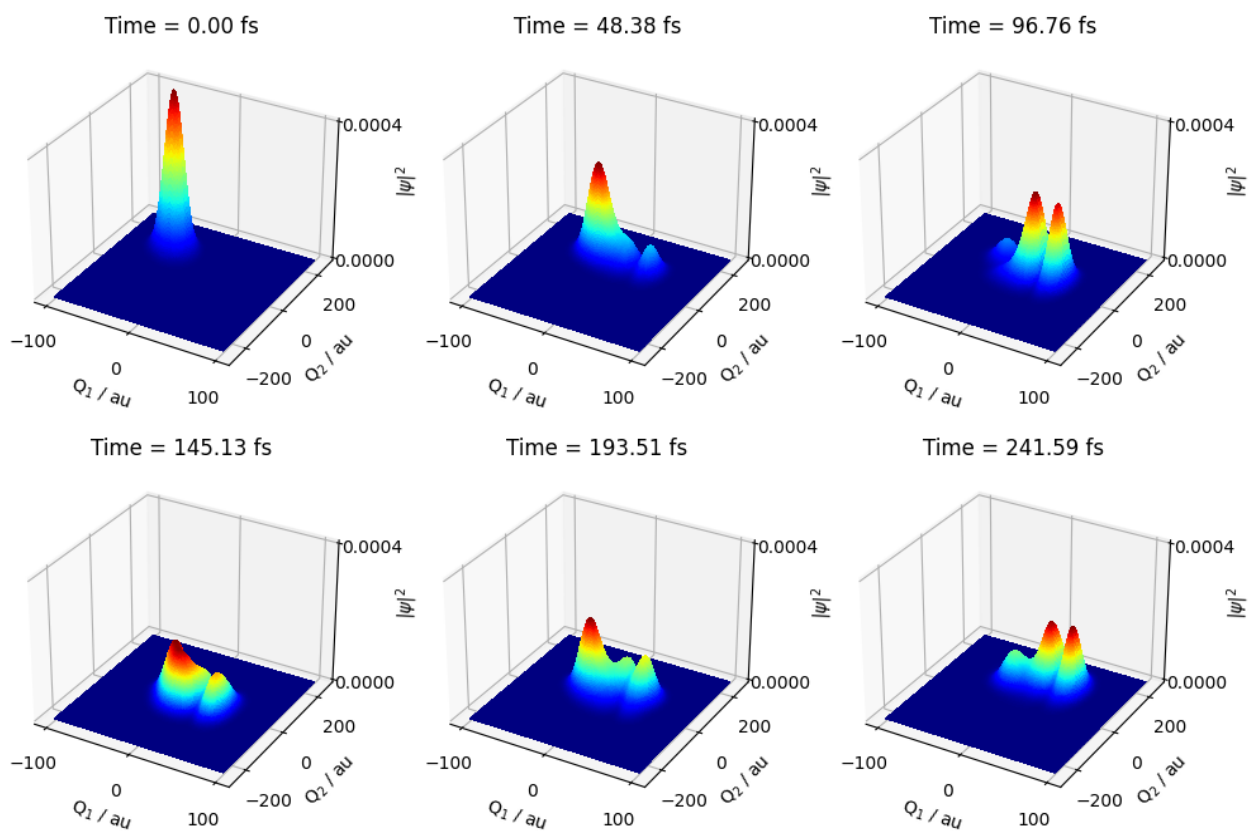

**Figure S8.** Probability density as a function of the large-amplitude mode coordinates $Q_1$ and $Q_2$ with integration over the 18 lowest frequency bath modes.

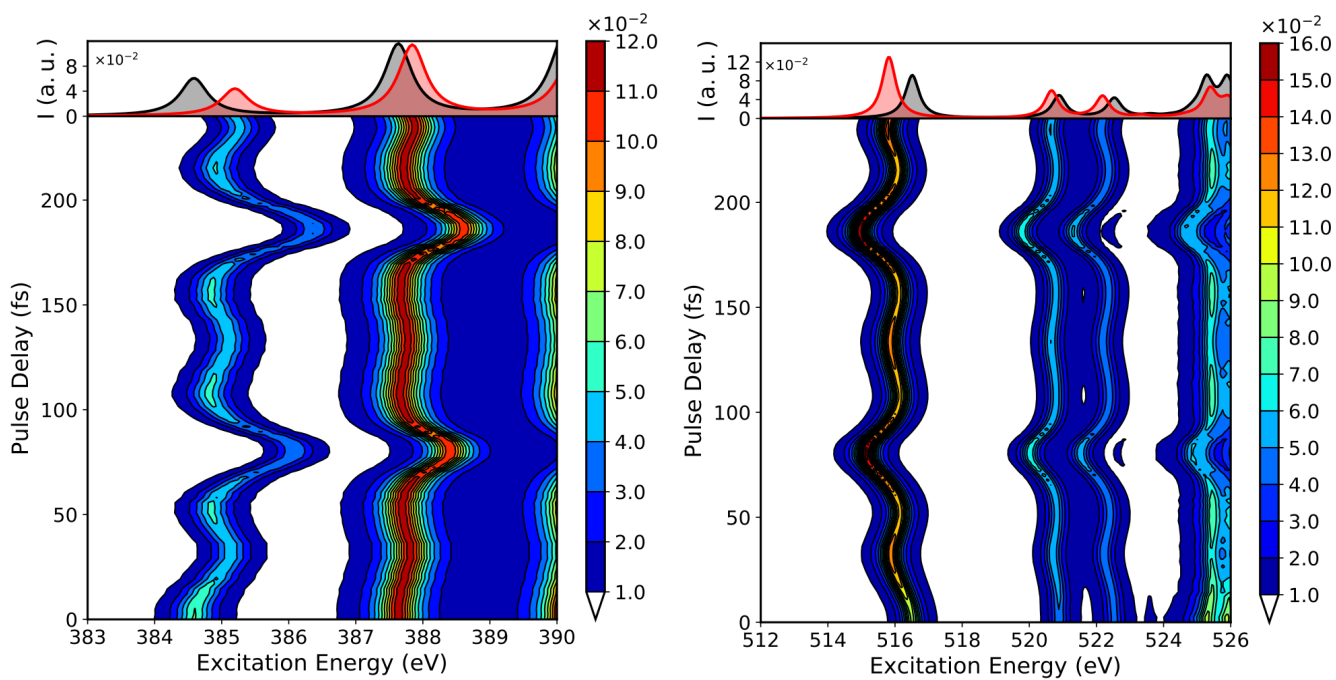

**Figure S9.** N K-edge (left) and O K-edge (right) TRXAS for HBT after photoexcitation with consideration of only the large-amplitude modes and omitting bath modes. Inset panel shows NEXAFS spectra at the initial enol* (black curve) and final keto* (red curve) conformation.

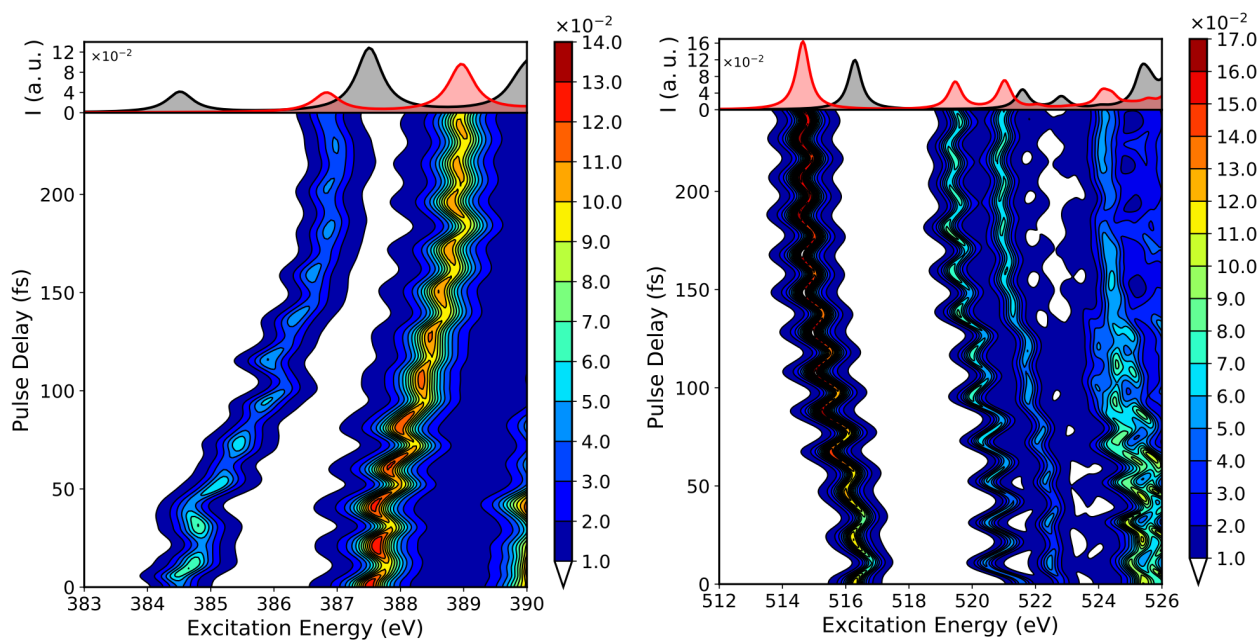

**Figure S10.** N K-edge (left) and O K-edge (right) TRXAS for HBT after photoexcitation considering the average trajectory. Inset panel shows NEXAFS spectra at the initial enol* (black curve) and final keto* (red curve) conformation.